\documentstyle{article}

\begin{document}
\begin{center}
{\LARGE Comment on ``A simple expression for the terms in the
  Baker--Campbell--Hausdorff series''} \\ \bigskip
{\large Hiroto Kobayashi} \\ \bigskip
{\it Department of Applied Physics, School of Engineering, 
     The University of Tokyo} \\
{\it Hongo, Bunkyo-ku, Tokyo 113-8656, Japan} \\
\end{center}

\begin{abstract}
It is pointed out that Reinsch's matrix operation formulation of
calculating the Baker--Campbell--Hausdorff series [math-ph/9905012] is
equivalent to the straightforward series expansion. The amount of
calculation does not decrease by his method.
\end{abstract}

Using matrix operations, Reinsch [1] proposed a simple expression of
terms in the Baker--Campbell--Hausdorff series of $ \log (\exp x \exp
y) $, where $ x $ and $ y $ are noncommutative variables. We here
point out that his formulation applies to the series expansion of
general functions of noncommutative variables. In fact, the matrix
formulation is equivalent to the straightforward series expansion in
the sense that the amount of calculation does not decrease by his
method. 

Let us show in the following the relation between the series expansion 
and the matrix operation. Suppose that $ a $ and $ b $ are polynomials
of noncommutative variables $ x, \  y, \  z, \ldots $. We express the
polynomials as 
\begin{equation}
a = \sum_{i=0}^\infty a_i, \quad b = \sum_{i=0}^\infty b_i ,
\end{equation}
where $ a_i $ and $ b_i $ denote the $ i $th-order terms. When we need 
to the series expansion of the product $ c = ab $ up to the $ n $th
order, we define two $ (n+1) \times (n+1) $ matrices $ A $ and $ B $
as 
\begin{equation}
A_{ij} = \sum_{k=0}^n a_k \delta_{i+k,j} , \quad
B_{ij} = \sum_{k=0}^n b_k \delta_{i+k,j} .
\end{equation}
We obtain the $ i $th-order term $ c_i $ of the product of $ a $ and $ 
b $ as 
\begin{equation}
c_i = C_{1,i+1} \quad {\rm with} \quad C = A B \quad {\rm for} \quad i 
\le n . 
\label{eq:c=ab}
\end{equation}

We derive Eq.\ (\ref{eq:c=ab}) in the following way. First we note
that 
\begin{eqnarray}
a b & = & ( a_0 + a_1 + \cdots + a_n + \cdots )
          ( b_0 + b_1 + \cdots + b_n + \cdots )
\nonumber \\
    & = & a_0 b_0 + ( a_0 b_1 + a_1 b_0 ) + \cdots 
      + ( a_0 b_n + a_1 b_{n-1} + \cdots + a_n b_0 ) + \cdots
\nonumber \\
    & = & \sum_{i=0}^\infty \sum_{j=0}^i a_j b_{i-j} , 
\end{eqnarray}
that is, 
\begin{equation}
c_i = \sum_{j=0}^i a_j b_{i-j} .  \label{eq:ci}
\end{equation}
On the other hand, the equation $ C = AB $ is 
\begin{equation}
\left(\begin{array}{ccccc}
c_0 & c_1 & \cdots & \cdots & c_n    \\
    & c_0 & c_1    &        & \vdots \\
    &     & \ddots & \ddots & \vdots \\
    &     &        & \ddots & c_1    \\
    &     &        &        & c_0
\end{array} \right)
=
\left(\begin{array}{ccccc}
a_0 & a_1 & \cdots & \cdots & a_n    \\
    & a_0 & a_1    &        & \vdots \\
    &     & \ddots & \ddots & \vdots \\
    &     &        & \ddots & a_1    \\
    &     &        &        & a_0
\end{array} \right)
\left(\begin{array}{ccccc}
b_0 & b_1 & \cdots & \cdots & b_n    \\
    & b_0 & b_1    &        & \vdots \\
    &     & \ddots & \ddots & \vdots \\
    &     &        & \ddots & b_1    \\
    &     &        &        & b_0
\end{array} \right) . \label{eq:C=AB}
\end{equation}
It is obvious that Eq.\ (\ref{eq:C=AB}) yields Eq.\ (\ref{eq:ci}), and
hence Eq.\ (\ref{eq:c=ab}) is proved. 

Furthermore, the $ i $th-order term of the summation $ a + b $ is
obtained from $ (A+B)_{1,i+1} $. Combining this with 
Eq.\ (\ref{eq:c=ab}), we obtain the fact that for a general polynomial
$ f(a) $, the $ i $th-order term $ [f(a)]_i $ is given by $
[f(A)]_{1,i+1} $. Moreover, the matrix $ f(A) $ has the same form of $
A $, that is 
\begin{equation}
[f(A)]_{ij} = \sum_{k=0}^n [f(a)]_k \delta_{i+k,j} .
\end{equation}
Thus for two general polynomials $ f(a) $ and $ g(b) $, the $ i
$th-order term $ [f(a) g(b)]_i $ is given by $ [f(A) g(B)]_{1,i+1}
$. A function $ h( f(a) g(b) ) $ is also calculated from $ h( f(A)
g(B) ) $. In the above way, we find that the matrix operation is
equivalent to the straightforward series expansion. 

Reinsch [1] specifically calculated $ \log (\exp M \exp N) $ for two $
(n+1) \times (n+1) $ matrices $ M $ and $ N $ defined as 
\begin{equation}
M_{ij} = \delta_{i+1,j} x =
\left( \begin{array}{ccccc}
0 & x & 0 & \cdots & 0      \\
  & 0 & x & \ddots & \vdots \\
  &   & 0 & \ddots & 0      \\
  &   &   & \ddots & x      \\
  &   &   &        & 0
\end{array} \right), \quad  N_{ij} = \delta_{i+1,j} y =
\left( \begin{array}{ccccc}
0 & y & 0 & \cdots & 0      \\
  & 0 & y & \ddots & \vdots \\
  &   & 0 & \ddots & 0      \\
  &   &   & \ddots & y      \\
  &   &   &        & 0
\end{array} \right) . 
\end{equation}
As shown above, it is only a special case that $ [\log (\exp M \exp
N)]_{1,i+1} $ gives the $ i $th-order term of $ \log (\exp x \exp y)
$. (Note that the logarithm and the exponential function of
noncommutative variables are defined by their polynomial expansion.)
The amount of calculation to obtain the $ i $th-order term does not
decrease by Reinsch's method. For example, in order to calculate the
product of $ \exp x $ and $ \exp y $, we need to multiply matrix
elements once for the zeroth order, twice for the first order, and
three times for the second order, as shown in the following equation:
\begin{equation}
\left( \begin{array}{ccc}
1 & x & \frac{1}{2} x^2 \\
0 & 1 & x               \\
0 & 0 & 1               
\end{array} \right)
\left( \begin{array}{ccc}
1 & y & \frac{1}{2} y^2 \\
0 & 1 & y               \\
0 & 0 & 1               
\end{array} \right) =
\left( \begin{array}{ccc}
1 & x+y & \frac{1}{2} x^2 + xy + \frac{1}{2} y^2 \\
0 & 1   & x+y                                    \\
0 & 0   & 1               
\end{array} \right) . 
\end{equation}
The number of multiplication is the same as the corresponding
straightforward series expansion. 

Finally, we would like to draw attention to the NCAlgebra package for
Mathematica [2], which is useful for series expansions containing
noncommutative variables. 

I would like to thank Prof.\ N.\ Hatano for his useful comments.

\bigskip
\noindent
[1] M.W.\ Reinsch: math-ph/9905012. 

\noindent
[2] \verb+http://math.ucsd.edu/~ncalg/+ .

\end{document}